\documentclass[12pt]{iopart}
\eqnobysec
\usepackage{graphicx}
\usepackage{color}

\font\amsb=msbm10
\def\hbar{\mbox{\amsb\char'175}}

\newcommand{\be}{\begin{equation}}
\newcommand{\ee}{\end{equation}}

\newcommand{\vecp}{{\mathbf p}}
\newcommand{\vecq}{{\mathbf q}}

\newcommand{\x}{{\mathbf x}}

\newcommand{\J}{{\mathbf J}}

\newcommand{\C}{{\mathbf C}}

\newcommand{\opA}{{\widehat{A}}}

\newcommand{\opR}{{\widehat{R}}}
\newcommand{\opT}{{\widehat{T}}}
\newcommand{\oprho}{{\widehat{\rho}}}

\newcommand{\opU}{\widehat{U}}
\newcommand{\opI}{\widehat{I}}

\newcommand{\oq}{\widehat{q}}
\newcommand{\op}{\widehat{p}}

\newcommand{\M}{{\mathbf M}}

\newcommand{\Id}{{\mathbf I}}

\newcommand{\der}{\partial}

\newcommand{\vct}[1]{\ensuremath\mbox{\boldmath$ #1 $}}
\newcommand{\Vxi}{{\vct \xi}}

\begin{document}

\title{Semiclassical evolution of correlations between observables}

\author {Alfredo M. Ozorio de Almeida\footnote{ozorio@cbpf.br}}
\address{Centro Brasileiro de Pesquisas Fisicas,
Rua Xavier Sigaud 150, 22290-180, Rio de Janeiro, R.J., Brazil}

\author{Olivier Brodier}
\address{Laboratoire de Math\'ematiques et Physique Th\'eorique, Faculte des Sciences et Techniques, 
Universit\'e de Tours, 37200 Tours, France}

\begin{abstract}

The trace of an arbitrary product of quantum operators with the density operator is rendered
as a multiple phase space integral of the product of their Weyl symbols with the Wigner function.
Interspersing the factors with various evolution operators, one obtains an evolving correlation.
The kernel for the matching multiple integral that evolves within the Weyl representation 
is identified with the trace of a single compound unitary operator. 
Its evaluation within a semiclassical approximation 
then becomes a sum over the periodic trajectories of the corresponding classical
compound canonical transformation. 

The search for periodic trajectories can be bypassed by an 
exactly equivalent initial value scheme, which involves a change of integration variable and
a reduced compound unitary operator. Restriction of all the operators to observables with smooth non-oscillatory Weyl symbols reduces the evolving correlation to a single phase space integral.
If each observable undergoes independent Heisenberg evolution, the overall correlation evolves 
classically. Otherwise, the kernel acquires a nonclassical phase factor, though
it still depends on a purely classical compound trajectory: e.g. the fase for a double return
of the quantum Loschmidt echo does not coincide with twice the phase for a single echo.

\end{abstract}

\maketitle

\section{Introduction}

Notwithstanding the aptness of semiclassical (SC) approximations for uncovering
classical structures underlying quantum evolution, 
their use for constructing ready algorithms to deal with increasingly complex experiments 
is yet to be established. The practical advantage of integrating Hamilton's ordinary differential equations, rather than dealing with Schr\"odinger's partial differential equation,
is counterbalanced by the need to search for trajectories that are only indirectly
specified by boundary conditions, instead of arising directly from their initial conditions.
This difficulty has led to the development of initial value methods (or {\it inital value representations}, IVR) that substitute the, so called, {\it root search} by an integration
over families of initial valued trajectories \cite{Mil70,Mil01,Mil02,FGrossmann,HerKLuck,Kay}.
Thus, IVR's have been seen as a workable alternative, 
in spite of considerable criticism \cite{BarAguiar01}. 
One of their main achievements is the evaluation of {\it correlations} for quantum operators,  ${\rm tr} \hat{A} \hat{B}(t)$ \cite{Mil12}, even though, if $\hat{A}$ is chosen as a density operator, 
this reduces to a single evolving expectation value $\langle \hat{B}(t) \rangle$.
Here we establish general SC approximations for multiple correlations of observables evolved 
by various unitary operators: $\langle\opU \hat{B} \hat{V} \hat{C}... \rangle$.

In a recent paper \cite{IVRFVR}, henceforth labled {\bf I}, the IVR approach to SC approximations 
was realized entirely within the Weyl representation, that represents the operator $\hat{B}$
by the phase space function $B(\x)=B(p,q)$, its {\it Weyl symbol}, 
or its Fourier conjugate, $\tilde{B}(\Vxi)$, its {\it chord symbol}, 
another complete representation \cite{Report}. 
A remarkable feature of these particular representations 
(including the Wigner function \cite{Wigner}, in the case of the density operator) 
is that they are based, respectively, on reflection operators and translation operators
\cite{Grossmann,Royer, Report}. These are themselves unitary, 
so one can combine them with the evolution operators which act on each observable
into a single composite entity. Then the expectation of an evolving observable was cast as a phase space integral over the Wigner function multiplied by the nearly classical function that represents the observable.

This procedure is here generalized to the correlations of an arbitrary number, $\nu$, of
observables undergoing general unitary evolutions. The evolving correlation depends 
on a single family of {\it compound unitary operators}, 
labeled by $\nu$ continuous parameters. For each parameter, the required trace of this compound operator
is then obtained from the {\it compound periodic orbits} in the corresponding classical evolution, 
according to the standard SC procedure \cite{Gutzwiller,livro}. 

Even though this is an important step, the identification of the appropriate compound unitary operator
does not free us from a search for orbits. It is true that continuous families of periodic orbits, 
within continuous families of canonical transformations, can be followed by a generalized Newton's method, as the parameters are varied in small steps \cite{Aguiaretal87}, 
but this is still a formidable task.
Furthermore, such a reliance on continuity is at odds with the use of efficient Monte Carlo methods 
at the next stage, where one integrates over the parameters. It is then fortunate that the IVR approach
can be extended to the general evaluation of correlations, by simply freeing one of the parameters: 
The corresponding segment of the periodic orbit is then removed, 
so that one then deals with a {\it reduced compound trajectory}. 
This is still composed of evolutions intercalated by reflections,  
but now the trajectory is determined by its initial value. There always exists an extra reflection 
which closes such an open orbit, so that its reflection centre can be chosen as the extra free parameter.

Just as in {\bf I}, the IVR algorithm avoids caustic singularities,
transforming them into nodal lines (or surfaces) of the compound propagator. There remains an overall
ambiguity of sign to be determined as such a line is crossed, but the general procedure
presented in \cite{OAI}, henceforth labled {\bf II}, can be immediately generalized for correlations. 

In the special case where all the evolution operators are {\it metaplectic}
\footnote{Unitary operators corresponding to classical symplectic transformations,
that is, linear canonical transformations, e.g. those driven by harmonic oscillators 
\cite{Peres, Bargmann, KramMoshSel, GuilStern, Voros76, Voros77, Littlejohn86, deGosson06}.},
the semiclassical theory is exact, including its IVR version. This provides scope
for simple applications that illustrate the general features of the method, without
gripping with the difficulties of a full SC calculation, as presented in {\bf I}. 
Perhaps, the greatest simplification concerns caustics (or nodal lines in IVR): 
The important point is that, even though families of metaplectic propagators 
do cross caustics in any representation, as a parameter is varied, 
in their case, the final integral for the correlation has no risk of
being divided into regions with different signs that need to be determined, 
as was discussed in {\bf II}.

The present paper follows closely on the track of {\bf I}: 
The same notations are adopted and we incorporate here many relevant features.
For instance, descriptions of any of the observables to be averaged may be supplied
by the translation operators underlying the chord representation, instead of the reflections
that belong to the Weyl representation, so that here we just focus on the latter. 
Again, we shall not develop explicitly the alternative of picking a pair of trajectories 
(forming a Final Value Representation, FVR)
the possible advantages being discussed in {\bf I}.
These alternatives shall remain implicit  
so as to emphasize the purely original features of the present work.
We shall also rely on the discussion of sign ambiguities associated with crossings
caustics in {\bf II}, since they are readily incorporated into the wider setting 
of evolving correlations. The great simplification here is one of scale:
By focusing on {\it mechanical observables} for which the Weyl symbol
coincides with the corresponding smooth classical phase space function except for corrections
that are of first order in $\hbar$, one reduces the expression for the evolving correlation
to a single phase space integral, irrespective of the number of observables.

The following section presents the general construction of the appropriate compound unitary operator
as the kernel for the correlation of evolving operators. Section 3 then interprets its 
SC approximation in terms of a compound canonical transformation defined by a sequence of trajectory segments and derives its trace from the periodic orbits. The alternative IVR scenario is then
developed in section 4, whereas section 5 presents the simplifications inherent to the propagation
of mechanical observables. All formulae are valid for an even number of observables. Modifications
that may be required in the odd case are discussed in the appendix.

\section{Compound unitary operators}

The outcome of a standard repeated experiment on a quantum system
is expressed as an average over an (observable) operator,  
which may correspond to a standard classical variable, such as position,
a projector, or a POVM. One can also measure correlations between such
observables that have undergone different evolutions. 
In the simplest case, these may concern the same operator
traversing coherently the alternative paths of an interferometer, or just
measured at different times, as in the correlations of Leggett-Garg \cite{Leggett-Garg}.
Then, so that the correlation is real, one evaluates some suitable symmetrization of
\begin{equation}
\C = \langle\opA_{\nu}(t_{\nu})...~\opA_2(t_2)\opA_1(t_1)\rangle 
= {\rm tr}~\opA_{\nu}(t_{\nu})~...~\opA_2(t_2)\opA_1(t_1)~\oprho,
\label{C1}
\end{equation}
where each of the Hermitian operators $\opA_j(t_j)$ undergoes a Heisenberg evolution
driven by some unitary operator, ${\widehat V}_j$, that is
\begin{equation}
\opA_j(t_j) = {{\widehat V}_j(t_j)}^{\dagger} \;\opA_j \;\; {\widehat V}_j(t_j).
\label{Heisenberg}
\end{equation}
If one defines the intermediate steps as
\begin{equation}
\opU_{j+1} \equiv{\widehat V}_{j+1}(t_{j+1}){\widehat V}_j(t_j)^{\dagger}
\label{intermediate}
\end{equation}
(with $\opU_1 \equiv{\widehat V}_1$ and $\opU_{\nu+1} \equiv{\widehat V}_{\nu}(t_{\nu})^{\dagger}$),
the general form is obtained, 
\begin{equation}
\C =  {\rm tr}~\opU_{\nu+1}\opA_{\nu}~\opU_{\nu}~...~\opA_2~\opU_2~\opA_1~\opU_1~\oprho,
\label{C2}
\end{equation}
in terms of the original observables $\opA_j$. No longer is one limited to symmetric Heisenberg evolutions, being that each observable can now be sandwitched between arbitrary unitary operators $\opU_j$. Hence, \eref{C2} also includes evolutions such as the fidelity, i.e. the quantum Loschmidt echo \cite{Gorin, IVRFVR}.
\footnote{This is the case of a single observable $\opA_1 = \opI$.} 
An example of direct application of such a time evolved correlation arises 
in the theory for time-resolved electronic spectra, depending on the evolution of two pairs of transition dipole operators. The Franck-Condon approximation then leads to an expression for
the spectrum in terms of the fidelity, which was obtained in \cite{ZamSVan} using IVR. 
The present theory supplies in principle the full correlation witout any supplementary approximation. 

Following the same notation as in {\bf I} and {\bf II},
the {\it centre symbol or Weyl symbol} of operator $\hat{A}$ is
\begin{equation}
\label{covW} 
A(\x) = 2^N{\rm tr}\;\left[\hat{R}_{\x}\;\hat{A}\right],
\label{Weylrep}
\end{equation}
where $N$ is the number of degrees of freedom, while the unitary operator, $\hat{R}_{\x}$,
corresponding to the (classical) reflection through the phase space point $\x$ 
\cite{Grossmann, Royer, Report}, plays a fundamental role throughout.
In other words, $A(\x)$ is the Weyl representation of $\hat{A}$.
Alternatively, the {\it chord symbol} of the operator $\hat{A}$ can be defined as
\begin{equation}
\label{covC} 
\tilde{A}(\Vxi) = {\rm tr} \;\left[\hat{T}_{-\Vxi}\;\hat{A}\right],
\end{equation}
where $\hat{T}_{\Vxi}$ is the {\it Heisenberg operator} corresponding to a phase space translation 
by the vector $\Vxi$. The chord symbol and the Weyl symbol 
are related by Fourier transformation. The advantage of both these representations is that 
their families of {\it basis operators},
$\{\hat{R}_{\x}\}$ and $\{\hat{T}_{\Vxi}\}$, belong to the group of unitary operators.
An arbitrary operator is then expressed as a superposition of unitary operators
\begin{equation}
\label{conW} 
\hat{A} = 2^N\int \frac{d\x}{(2\pi\hbar)^N}\; A(\x) \;
\hat{R}_{\x} ~
= \frac{1}{(2\pi\hbar)^N}\int d\Vxi \;
\tilde{A}(\Vxi)\; \hat{T}_{\Vxi} \,
\end{equation}
though it is convenient to keep the special notation for the density operator,
\begin{equation}
\label{rho} 
\oprho = 2^N\int d\x\; W(\x) \;
\hat{R}_{\x} ~
= \int d\Vxi \; \chi(\Vxi)\; \hat{T}_{\Vxi} \,
\end{equation}
in terms of the Wigner function \cite{Wigner}, $W(\x)$, 
and the chord function \cite{Report}, $\chi (\Vxi)$.

In the case of the Weyl-Wigner representation, one can now insert these 
in the expression for the evolving correlation
\begin{equation}
\fl \C = \frac{2^N}{(\pi\hbar)^{\nu N}}\int d\x_{\nu}... d\x_2d\x_1d\x_0\; A_{\nu}(\x_{\nu})...A_2(\x_2)~A_1(\x_1)~W(\x_0)~~
{\rm tr}~ {\widehat{\mathbf U}}\{\x_0,\x_1,\x_2,...,\x_{\nu}\},
\label{C3}
\end{equation}
where the family of {\it compound unitary operators} is defined as
\begin{equation}
{\widehat{\mathbf U}}\{\x_0,\x_1,\x_2,...,\x_{\nu}\} 
\equiv \opU_{\nu+1}~\hat{R}_{\x_{\nu}}~\opU_{\nu}...~\hat{R}_{\x_1}\opU_1~\hat{R}_{\x_0}.
\label{compu1}
\end{equation}

Before any evolution takes place, that is, when all of the $t_j=0$ in \eref{C1},
the compound operator is just a product of reflections, because all the $\opU_j=\opI$, 
the identity operator:
\be
{\widehat{\mathbf U}}_0\{\x_0,\x_1,\x_2,...,\x_{\nu}\}
=\hat{R}_{\x_{\nu}}~...~\hat{R}_{\x_1}~\hat{R}_{\x_0}.
\label{compu0}
\ee
The simplest case is when $\nu$, the number of reflections is odd, i.e. an even number
of observables. Then the product is also a reflection and we can identify
${\rm tr}~ {\widehat{\mathbf U}}_0$ with the kernel for the Weyl representation 
of a product of Weyl operators, each of which is specified by its Weyl symbol \cite{Report}:
\be
{\rm tr}~ {\widehat{\mathbf U}}_0\{\x_0,\x_1,\x_2,...,\x_{\nu}\}
= 2^{-\nu N} \exp\left[\frac{i}{\hbar} \Delta_{\nu+1}(\x_0,\x_1,\x_2,...,\x_{\nu})\right],
\label{tr0}
\ee 
where $\Delta_{\nu+1}$ is the symplectic area of the polygon whose sides are centred on $\{\x_0,\x_1,\x_2,...,\x_{\nu}\}$ as drawn in Fig 1; this is a bilinear function of each pair of variables, which arises in the general product formula for the Weyl representation 
of the product of an even number of operators, $\opA_{\nu}...\opA_1$, in \cite{Report}, that is,
\be
\fl \{A_{\nu}...A_1\}(\x_0) = \int \frac{d\x_{\nu}...d\x_1}{(\pi\hbar)^{\nu N}}~ A_{\nu}(\x_{\nu})...A_1(\x_1)~
\exp\left[\frac{i}{\hbar}\Delta_{\nu+1}(\x_0,\x_1,...,\x_{\nu})\right],
\label{Weylproduct}
\ee
so that one retrieves the simple expression for the initial correlation:
\be
\C_0 = \int d\x_0~ \{A_{\nu}...A_1\}(\x_0)~ W(\x_0).
\ee

In the following section, we show how this {\it polygonal scenario} is extended
to evolving correlations within the SC approximation.
\begin{figure}[htb!]
\centering
\includegraphics[height=4cm]{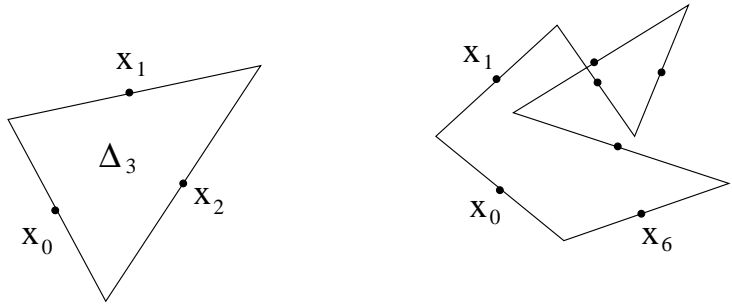}
\caption{The initial kernel of the integral for the correlation of an even number of observables
is the complex exponential of $\Delta_{\nu+1}(\x_0,\x_1,\x_2,...,\x_{\nu})$, the symplectic area of the unique polygon with sides centred on $\{\x_0,\x_1,\x_2,...,\x_{\nu}\}$. 
For, $\nu=2$, this is a triangle (left), whereas the plane projection of higher polygons may have self-intersections (right).}
\label{Figt0}
\end{figure}
Evidently, one can express evolving correlations equally well in terms of compound operators
with translations in place of reflections,
\begin{equation}
{\widehat{\mathbf U}}\{\Vxi_0,\Vxi_1,\Vxi_2,...,\Vxi_{\nu}\} 
\equiv \opU_{\nu+1}~\hat{T}_{\Vxi_{\nu}}~\opU_{\nu}...~\hat{T}_{\Vxi_1}\opU_1~\hat{T}_{\Vxi_0},
\label{compu2}
\end{equation}
or it may be more convenient to keep some reflections and some translations.
The issue appears in the context of a single evolving expectation and it is already
discussed in {\bf I}. Indeed, the product of any even number of reflections, 
${\widehat{\mathbf U}}_0\{\x_0,\x_1,\x_2,...,\x_{\nu}\}$, results 
in a translation operator, rather than a reflection, so that it is advantageous
to also use the chord-translation basis for the density operator.
For this reason, the results that must be adapted for an odd number of observables, 
by resorting to the Fourier transform of the Wigner-Weyl representation, 
will be remitted to the appendix.

\section{SC approximation for the compound Weyl propagator}

The key ingredient for the SC approximation of the compound propagator and its trace
is the general SC {\it Weyl propagator} corresponding to an arbitrary 
classical canonical transformation that is
generated by a Hamiltonian, $H(\x)$, acting during a time, $t$. 
In the simplest case, this is simply \cite{Berry89, Report}
\begin{equation}
U(\x) \approx \frac{2^N}{|\det(\Id+ \M)|^{1/2}}\>\>
\exp \left[ \frac{i}{\hbar}(S(\x)+ \hbar \pi \sigma)\right].
\label{Uweyl}
\end{equation}
The geometric part of the centre or Weyl action, $S(\x)$, is just the symplectic area between
the trajectory and the chord, $\Vxi={\x}^+ -{\x}^-$, joining its endpoints.
From this, one subtracts $-Et$, where $E$ is the energy of the trajectory.
The {\it Maslov index}, $\sigma$, is zero in a neighbourhood
of the identity operator, in which case there is indeed only a single classical trajectory 
centred on the point $\{\x=(\vecp, \vecq)\}$ of classical phase space ${\bf R}^{2N}$.
Otherwise, there may be multiple solutions to the variational problem that identifies trajectories 
with a given centre, $\x$, so the actions may have many branches 
and these branches meet along  caustics where the semiclassical amplitude diverges. 
(See {\bf II} for the phase $\sigma$ and further details.)

The centre action specifies the classical canonical transformation, ${\x}^- \mapsto {\x}^+$,
corresponding to $\opU$ indirectly through \cite{Report}
\begin{equation}
\Vxi = -\J \frac{\der S}{\der \x}, \>\> {\x}^+ = \x + \frac{\Vxi}{2},
\>\> \x^-  = \x - \frac{\Vxi}{2}.
\label{centran}
\end{equation}
The linear approximation of this transformation near the $\x$-centred trajectory is defined
by the {\it symplectic} matrix $\M$. This has the Cayley parametrisation:
\begin{equation}
\M = [\Id + \J\mathbf{B}]^{-1} [\Id - \J\mathbf{B}],
\label{Cayley}
\end{equation}
in terms of the symmetric matrix $\mathbf{B}$, which is just the Hessian matrix of $S(\x)$.

A notable exception for this SC form of the Weyl propagator is precisely 
that of a reflection operator, $\opR_\x$. Indeed, its Weyl symbol is simply,
\be
R_\x(\x') = 2^{-N} \delta(\x'-\x).
\ee
It is the chord representation of this operator that has the standard semiclassical form. 

To construct the SC approximation of the compound propagator, 
${\widehat{\mathbf U}}\{\x_0,\x_1,\x_2,...,\x_{\nu}\}$, we assume that each of the Weyl propagators, ie. the Weyl symbols, $U_j(\x)$, for ${\opU}_j$ can be expressed in the form \eref{Uweyl}.  
Then the key point is that the compound unitary operator, 
${\widehat{\mathbf U}}\{\x_0,\x_1,\x_2,...,\x_{\nu}\}$, 
for any choice of parameters, has its own Weyl symbol, 
${\mathbf U}\{\x_0,\x_1,\x_2,...,\x_{\nu}\}(\x)$ or ${\mathbf U}(\x)$ for short.  
This compound propagator is determined from the factor propagators, $U_j(\x'_j)$ 
according to the product formula \eref{Weylproduct}:
\begin{eqnarray}
\fl {\mathbf U}\{\x_0,\x_1,\x_2,...,\x_{\nu}\}(\x)= 
\int \frac{d\x'_{\nu+1}d\x''_{\nu}...d\x'_1 \x''_0}{(\pi\hbar)^{2(\nu+1) N}} ~
&& U_{\nu+1}(\x'_{\nu+1})~R_{\x_{\nu}}(\x''_{\nu})...U_1(\x'_1)~ R_{\x_0}(\x''_0) \nonumber \\
&&\exp\left[\frac{i}{\hbar}\Delta_{2\nu+3}(\x''_0,\x'_1,...,\x''_{\nu}, \x'_{\nu+1})\right].
\label{compprop}
\end{eqnarray}
The integrals over the arguments of the reflections merely fix the respective centres, $\x''_j=\x_j$,
whereas all other integrals are evaluated semiclassically by stationary phase. 
But the deduction of \eref{Uweyl} in \cite{Report} was itself based on the same product formula,
that is, from a product of infinitesimal small time propagators, 
so the result is again a Weyl propagator of the same form, 
but it is constructed from a {\it compound trajectory} built up from the sequences
of partial trajectory segments.
The relevant trajectory, which is centred on $\x$, i.e. the argument of the Weyl propagator,
is built up out $\nu +1$ such segments joined by $\nu +1$ reflections.

\begin{figure}[htb!]
\centering
\includegraphics[height=4cm]{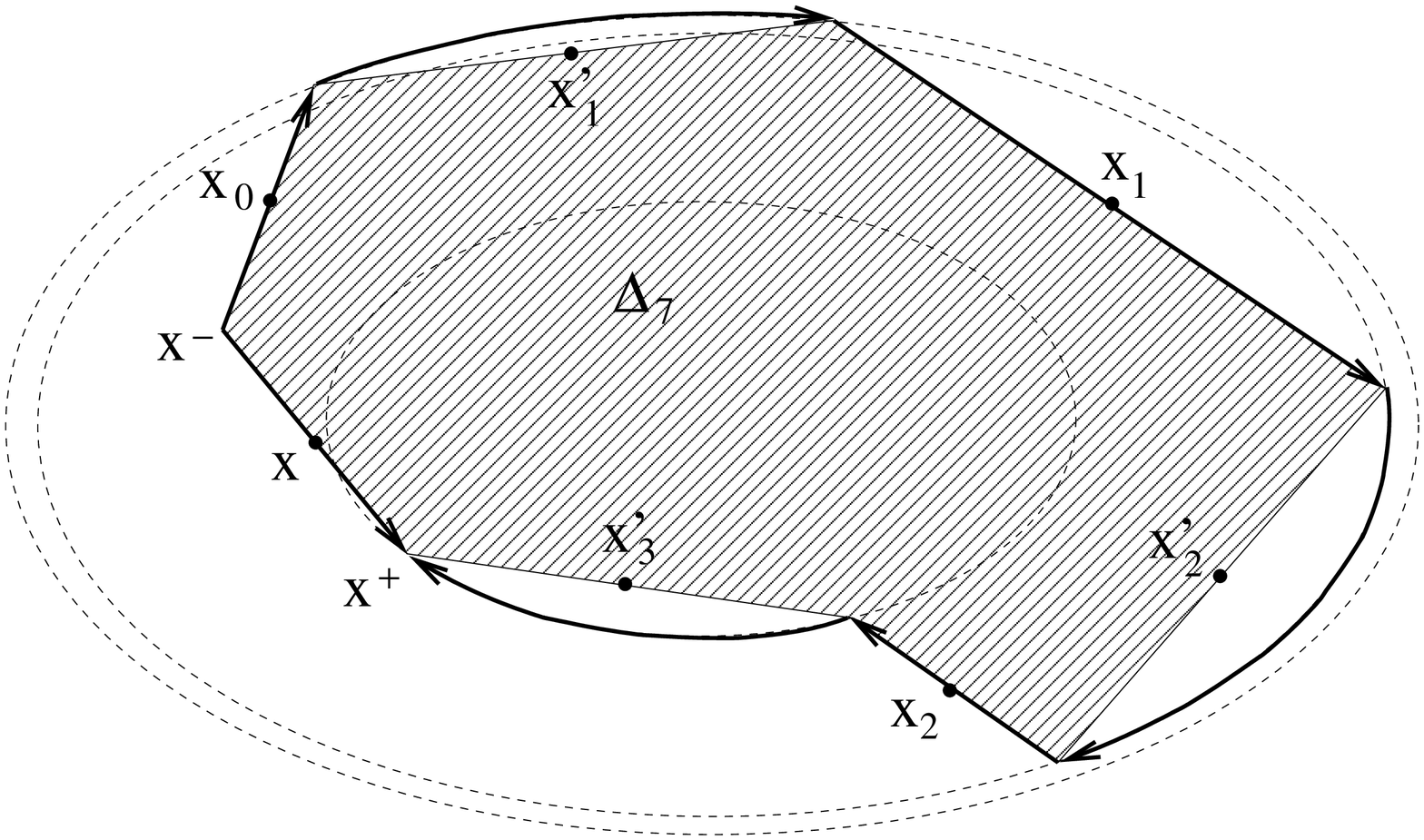}
\caption{ Phase space structure corresponding to the compound Weyl propagator for the evolution
of a pair of observables. The arguments of their Weyl symbol are the reflection centres
$\x_1$ and $\x_2$, while the reflection centre for the Wigner function is $\x_0$.
These reflections are joined by the trajectories of the intervening SC Weyl propagators
with centres at ${\x'}_1$ and ${\x'}_2$ so as to form the dynamical polygon. 
(These trajectories need not be generated by the same Hamiltonian)} 
\label{Figcom}
\end{figure}

The appropriate trajectory segments that are generated by each centre generating function,
$S_j({\x'}_j)$, as well the centres themselves, need to be chosen so that 
the full compound trajectory is continuous. This is achieved by imposing the requirement \cite{Report}
that the overall action in \eref{Uweyl} is just
\begin{equation}
{\mathbf S}(\x) = \Delta_{2\nu+3} + S_1({\x'}_1) +...+ S_{\nu+1}({\x'}_{\nu+1}).
\label{compaction}
\end{equation}
Here $\Delta_{2\nu+3}$ is the symplectic area of the {\it dynamical polygon} with a side centred on $\x$, as well as $\nu+1$ sides centred on the points $\x_j$ and $\nu+1$ sides centred on ${\x'}_j$. 
The stationary conditions,
\begin{equation}
\frac{\partial {\mathbf S}} {\partial{\x'}_j} = 0 ~~~ {\rm or} ~~~
\frac{d{\Delta}_{2\nu+2}}{{d\x'}_j} = -~\frac{dS_j}{{d\x'}_j}= {\Vxi}_j,
\label{statcon0}
\end{equation}
define the variables ${\x'}_j$, so that the trajectory arcs fit precisely the sides of
the dynamical polygon, as depicted in Fig. \ref{Figcom}, which exemplifies the general layout for the case of two observables, i.e. $\nu=2$. It is naturally satisfied by any compound trajectory 
that is followed through from its initial value.
As for the compound monodromy matrix, $\M$, this is just the product of the sequence of 
monodromy matrices for each step. Indeed, since the matrix for a reflection is just $-\Id$, 
we have simply
\begin{equation}
\M = [-\Id] \cdot \M_1 \cdot [-\Id] \cdot \M_2 \cdot ... [-\Id] \cdot \M_{\nu+1}
= (-1)^{\nu+1} \M_1 \cdot \M_2 \cdot ... \M_{\nu+1} .
\label{compmon}
\end{equation}

\begin{figure}[htb!]
\centering
\includegraphics[height=4cm]{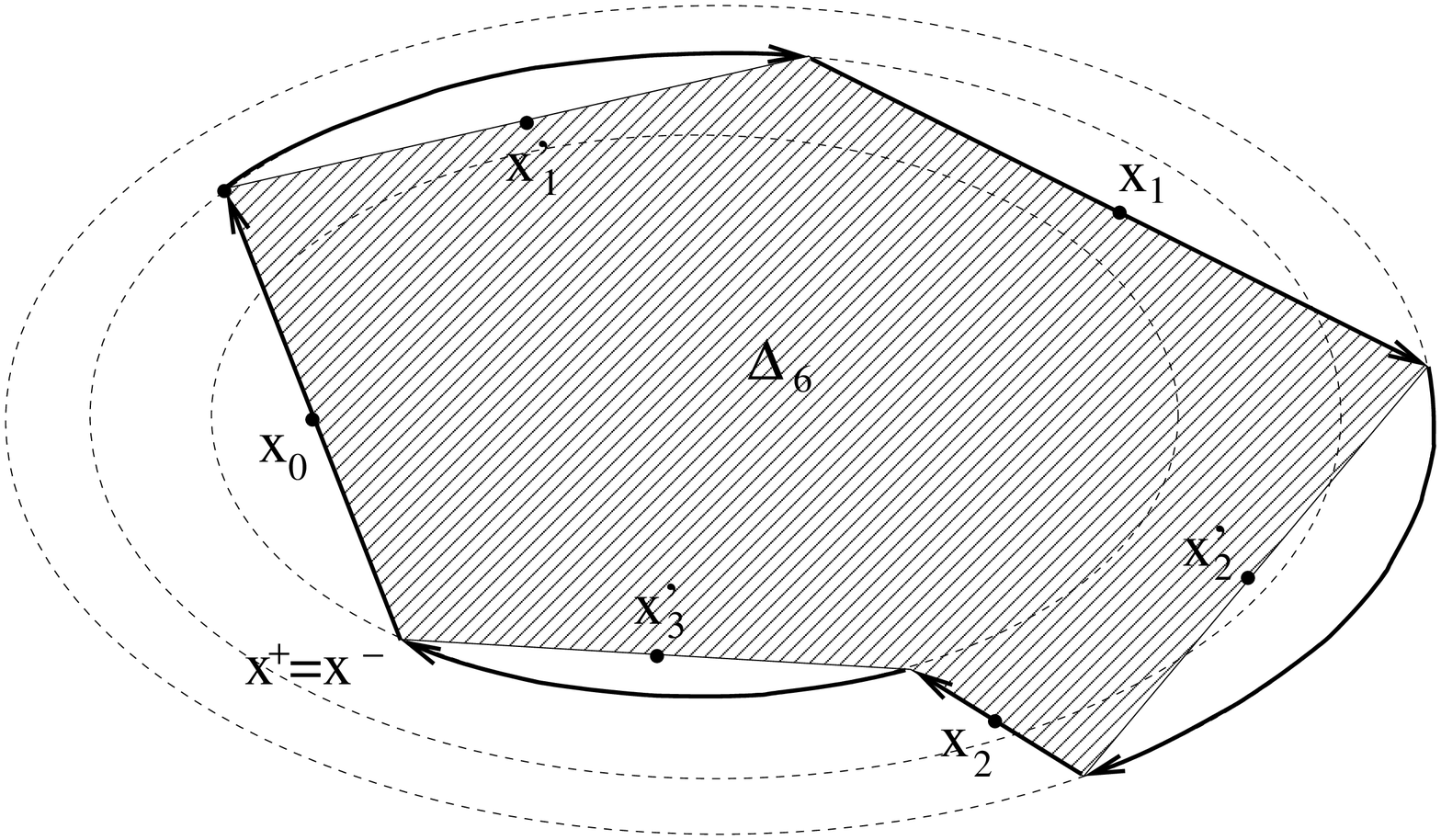}
\caption{ The SC expression for trace of the compound propagator is determined by
the corresponding periodic trajectories: The {\it open side} of the dynamical polygon, 
$\Delta_7$ in the previous figure collapses, so that $\x=\x^{-}=\x^{+}$.} 
\label{Figtr}
\end{figure}

The kernel for the propagation of the correlation between observables, ${\hat A}_j$,  
evaluated at multiple times in terms of the initial Weyl symbol, $A_j(\x)$,
is simply ${\rm tr}~{\widehat{\mathbf U}}\{\x_0,\x_1,\x_2,...,\x_{\nu}\}$.
Just as for other representations, the SC limit of the trace of an unitary operator
singles out the contributions of the periodic trajectories. In the case of the Weyl
symbol, this is derived from
\begin{equation}
{\rm tr}~ \widehat{{\mathbf U}} = \int \frac{d\x}{(2\pi\hbar)^N}~{\mathbf U}(\x).
\label{tru}
\end{equation}
Now, the only explicit dependence on the argument, $\x$ of the compound propagator
is in the symplectic area of the dynamical polygon, $\Delta_{2n+3} = \x \wedge \Vxi + Const$, 
where $\Vxi$ is the side centred on $\x$. Like the term $Const$, this depends only on the other centres \cite{Report}:
\begin{equation}
\frac{\Vxi}{2} = ({\x'}_1 - \x_0) + ({\x'}_2 - \x_1) + ...+({\x'}_{\nu+1} - \x_{\nu}).
\end{equation} 
Thus, 
\begin{equation}
\int \frac{d\x}{(2\pi\hbar)^N}~ \exp\left[\frac{i}{\hbar}\Delta_{2\nu+3}(\x, \x_j,{\x'}_j)\right] =
\exp\left[\frac{i}{\hbar}\Delta_{2\nu+2}(\x_j,{\x'}_j)\right] \delta(\Vxi).
\label{zerochord} 
\end{equation}
that is, the integral singles out those polygons where the open side has zero length. 
The compound action for the trace is the same as \eref{compaction}, 
but with $\Delta_{2\nu+3} \mapsto \Delta_{2\nu+2}$, which is the polygon for a periodic orbit. 
Such a {\it compound periodic orbit} for the trace is depicted in Fig.\ref{Figtr}.
The corresponding monodromy matrix is just \eref{compmon} without the last factor, $[-\Id]$. 

The stationary condition for the intermediate centres, ${\x'}_j$, of the periodic orbit
is just \eref{statcon0}, so that this merely prescribes each side of the dynamical polygon 
to be the chord for the corresponding trajectory segment. However, in practice 
one need not search for the solution of each of these equations
because they are automatically satisfied for any given periodic orbit of the compound
canonical transformation, i.e. one then has all the centres and the sides of the $(2\nu)$-sided
polygon for a particular pinning of the reflection centres $\{\x_0, \x_1, ..., \x_{\nu}\}$.
Then one needs only to evaluate \eref{tru} by stationary phase,
so that the contribution of the p'th periodic orbit to the trace is just
\begin{eqnarray}
{\rm tr}~ \widehat{\mathbf U}_p &&\approx \frac{2^N}{|\det {\mathbf B}\det(\Id+ \M)|^{1/2}}\>\>
\exp \left[ \frac{i}{\hbar}({\mathbf S}(\x)+ \frac{\hbar \pi \sigma'}{4})\right] \nonumber\\
&&= \frac{2^N}{|\det(\Id - \M)|^{1/2}}\>\>
\exp \left[ \frac{i}{\hbar}({\mathbf S}(\x)+ \frac{\hbar \pi \sigma'}{4})\right],
\end{eqnarray}
where, following \cite{Report}, one takes the determinant of \eref{Cayley}.
\footnote{The stationary phase evaluation adds a further phase of $\pi/4$ 
times the signature of $\mathbf B$. This will not be needed in the IVR theory in the next section,
so that it is here just included in the {\it Maslov index} $\sigma'$.} 
The parameters $\{\x_0,\x_1,\x_2,...,\x_{\nu}\}$ vary continuously 
in the ultimate integration for the correlations,
so that the periodic orbits at each neighbouring parameter can be found 
by a generalized Newton's method, such as in \cite{Aguiaretal87}. 

The amplitude of the periodic orbit contribution to 
${\rm tr}~ \widehat{\mathbf U}\{\x_0,\x_1,\x_2,...,\x_{\nu}\}$ becomes singular at those values of
$\{\x_0,\x_1,\x_2,...,\x_{\nu}\}$ that determine periodic orbit resonances: $\det(\Id - \M)=0$. This problem is avoided by the generalization of the IVR theory of {\bf I} in the following section. 
The IVR approach even dispenses with the search for periodic orbits 
in the evaluation of evolving correlations.

\section{Initial value representation}

Let us reinterpret the kernel of the evolving correlation \eref{C3} as
\begin{equation}
\fl {\rm tr}~ \widehat{\mathbf U}\{\x_0, \x_1, ..., \x_{\nu}\}
= {\rm tr}~ \opU_{\nu+1}~\hat{R}_{\x_{\nu}}~\opU_{\nu}...~\hat{R}_{\x_1}\opU_1~\hat{R}_{\x_0}~
= ~ {\mathbf U'}\{\x_1, ..., \x_{\nu}\}(\x_0),
\end{equation}
that is, according to \eref{Weylrep}, the Weyl symbol for the {\it reduced compound operator}:
\begin{equation}
\widehat{\mathbf U'}\{\x_1, ..., \x_{\nu}\} \equiv \opU_{\nu+1}~\hat{R}_{\x_{\nu}}~\opU_{\nu}...~\hat{R}_{\x_1}\opU_1.
\end{equation}
In its terms one generalizes the Weyl representation to the evolved product operator 
in \eref{Weylproduct} as
\be
\fl \{A_{\nu}...A_1\}'(\x_0) = \frac{2^N}{(\pi\hbar)^{\nu N}} 
\int d\x_{\nu}...d\x_1 ~ A_{\nu}(\x_{\nu})...A(\x_1) ~ 
{\mathbf U'}\{\x_1, ..., \x_{\nu}\}(\x_0),
\label{evproduct}
\ee
so that the correlation is simply
\be
\C = \int d\x_0 ~ W(\x_0) ~ \{A_{\nu}...A_1\}'(\x_0).
\label{Csimple}
\ee

\begin{figure}[htb!]
\centering
\includegraphics[height=4cm]{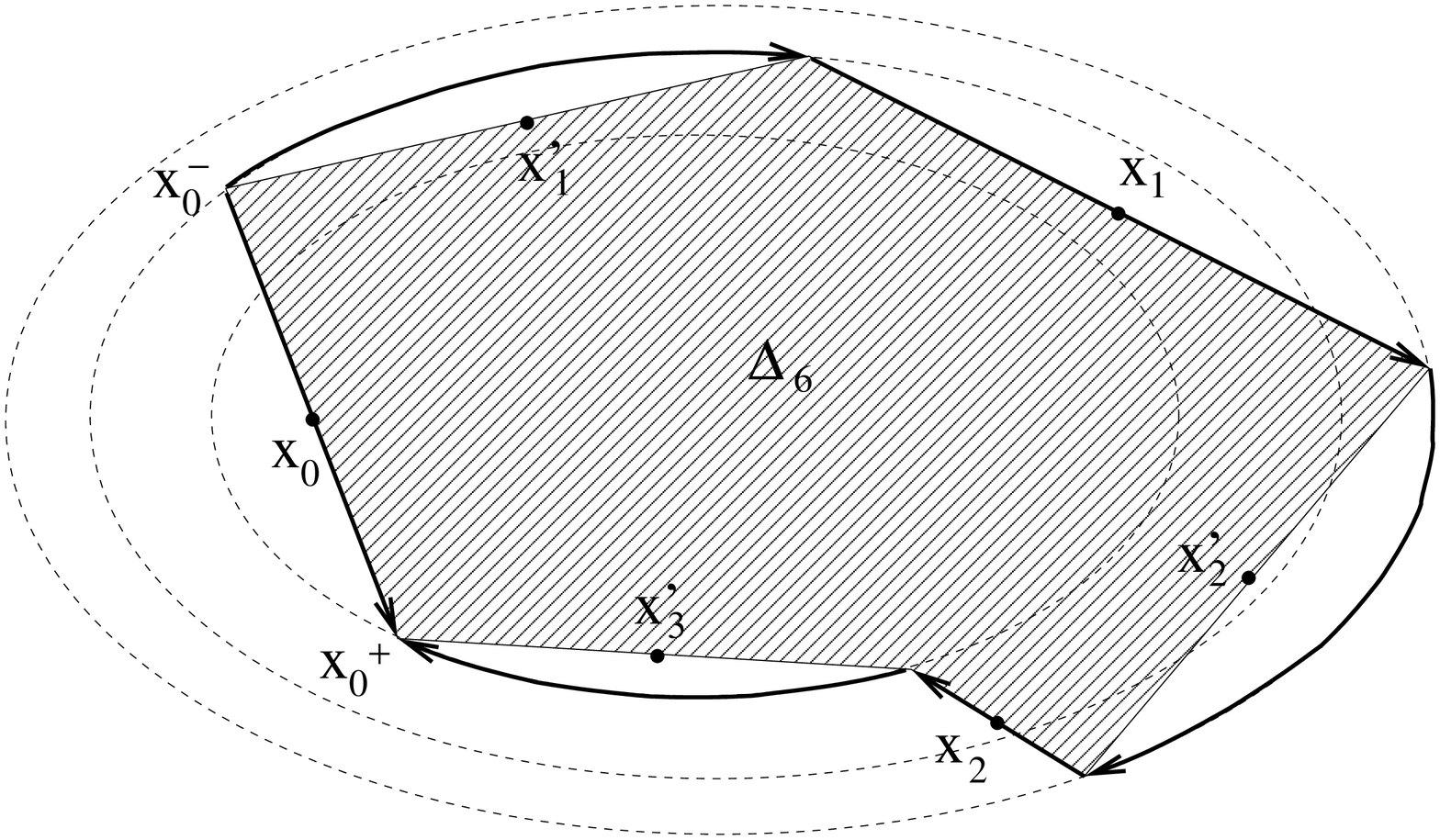}
\caption{ The same polygonal trajectory as in the previous figure is reinterpreted 
as an open trajectory of the {\it reduced} compound canonical transformation, from which
the initial reflection at $\x_0$ is removed. 
This point is now the argument of the Weyl symbol for the reduced compound propagator.}
\label{FigIVR}
\end{figure}

Thus, $\widehat{\mathbf U'}$ has one less reflection than $\widehat{\mathbf U}$, but clearly
its SC approximation has the same form as \eref{Uweyl}, with the corresponding 
classical polygonal path given by Fig.\ref{Figtr}, rather than Fig.\ref{Figcom} . 
In other words, Fig.\ref{Figtr} is now reinterpreted as an open polygonal line,
with its begining and end points, ${\x_0}^{\pm}$, centred on $\x_0$, 
as shown in Fig.\ref{FigIVR}.
It is important to note that, even though the full canonical transformation, 
${\x_0}^{-} \mapsto {\x_0}^{+}$, can be decomposed into several 
partial canonical transformations, the quantum unitary transformation $\widehat{\mathbf U'}$
corresponds to this single reduced compound canonical transformation. Furthermore,
each branch of its  centre generating function , ${\mathbf S'}(\x_0)$, in the SC approximation  
to the Weyl propagator, ${\mathbf U'}(\x_0)$, is constructed from those compound trajectories 
that satisfy ${\x_0}^{-} + {\x_0}^{+} = 2\x_0$.

The {\it initial value representation} (IVR) now results from the exchange of the integration variable
from the trajectory midpoint, $\x_0$,  that is the argument of the Wigner function in the evolving correlation \eref{C3}, to the initial point, ${\x_0}^{-}$. This generalization of the
procedure in {\bf I} relies on the simple form of the Jacobian of the transformation:
\be
\det \frac{d\x_0}{d{\x_0}^{-}} = \det \left( \frac{\Id+\M'}{2}\right).
\ee
Thus, it brings the SC approximation for the evolving correlation to the form
\begin{eqnarray}
\C \approx \frac{1}{(\pi\hbar)^{\nu N}}\int d\x_{\nu}...d\x_1 d{\x_0}^{-}                 A_{\nu}(\x_{\nu})...~A_1(\x_1)~W(\x_0({\x_0}^{-})) |\det(\Id + \M')|^{1/2} \nonumber \\
\exp\left[ \frac{i}{\hbar}({\mathbf S'}(\x_0({\x_0}^{-}))+ \hbar \pi \sigma)\right],
\label{C4}
\end{eqnarray}
where $\M'$ is the monodromy matrix for the linearization of 
${\x_0}^{-} \mapsto {\x_0}^{+}$ in the neighbourhood of ${\x_0}^{-}$.
It can be decomposed as the product 
\begin{equation}
\M' =  \M_1 \cdot [-\Id] \cdot \M_2 \cdot ... [-\Id] \cdot \M_{\nu+1},
\label{compmon2}
\end{equation}
a reduced version of \eref{compmon}, but now the sequence of factor monodromy matrices
is directly determined  by the initial value ${\x_0}^{-}$.
Likewise, each of the variables ${\x'}_j$ will be just the centre of of the respective
side ${\Vxi'}_j$ of the polygonal path starting at ${\x_0}^{-}$.
The crucial point is that the Jacobian for the switch to the new integration variable, 
${\x_0}^{-}$, kills off the caustic singularities in the SC kernel 
for the evolving correlation. Thus, in a single step, without increasing the number of integrations,
one does away with the need to search for trajectories while erasing all caustics! 

The generating function, ${\mathbf S'}(\x_0)$, for the canonical transformation
is now defined as
\begin{equation}
{\mathbf S'}(\x_0) = {\Delta'}_{2\nu+2}(\x_0, {\x'}_1,..., \x_{\nu}, {\x'}_{\nu+1}) 
+ S_1({\x'}_1) +...+ S_{\nu+1}({\x'}_{\nu+1}),
\label{compaction2}
\end{equation}
where the requirement 
\begin{equation}
\frac{\partial {\mathbf S'}} {\partial{\x'}_j} = 0 ~~~ {\rm or} ~~~
\frac{d{\Delta'}_{2\nu+2}}{{d\x'}_j} = -~\frac{dS_j}{{d\x'}_j}= {\Vxi'}_j,
\label{statcon}
\end{equation}
defining the variables ${\x'}_j$
is automatically satisfied by any compound trajectory that is followed through
from its initial value. Indeed, as discussed in \cite{Report}, 
the symplectic area of the {\it reduced dynamical polygon} satisfies
\begin{equation}
2 {\Delta'}_{2\nu+2} = {\Vxi'}_1 \wedge {\Vxi}_1 + ({\Vxi'}_1 + {\Vxi}_1)\wedge {\Vxi'}_2 + ... 
+ ({\Vxi'}_1 + {\Vxi}_1 + {\Vxi'}_2 + ... +{\Vxi'}_\nu)\wedge {\Vxi'}_{\nu+1}, 
\label{primepoly}
\end{equation}
where each chord refers to the appropriate reflection (unprimed) or partial evolution (primed).
The discussion of the Maslov phase $\sigma$ in {\bf II} applies direcly to \eref{C4}. 

In a full SC calculation in which each trajectory segment needs to be integrated numerically,
the numerical error will build up along the sequence of segments. It may then be preferable
to start somewhere in the middle of the sequence, from where the sequence is taken partly backwards
and partly forwards. This is a direct generalization of the {\it Final Value Representation} (FVR)
for the evolved average of a single observable that was presented in {\bf I}, even though
the forward and backward paths only have the same number of segments if $\nu$ is odd.
In any case, the Jacobian for the exchange of the initial value for an intermediate value,
${\x_0}^{-} \mapsto {\x_j}^{-}$, a canonical transformation, is just unity.

\section{Evolving mechanical observables}

So far no account has been taken of features that distinguish observables from other operators.
Within the Weyl representation, {\it mechanical observables} are represented by real smooth functions
of the phase space variables. Indeed, the Weyl representation of the operator function, 
$A(\hat{\x})$, is just the classical phase space function, $A(\x)$, plus corrections of order $\hbar$, which depend on the chosen symmetrization of products of $\op$ and $\oq$. Even more to the point, the Weyl representation of a product of mechanical observables, 
${\opA}_{\nu}...{\opA}_1$ is just $\{A_{\nu}...A_1\}(\x)=A_{\nu}(\x)...A_1(\x)$, up to first order terms in $\hbar$, 
which again depend upon ordering.

It is important to understand how this simplification arises, starting from \eref{Weylproduct}.
So one adapts the discussion concerning equations \eref{zerochord} and \eref{statcon}:
In the absence of any other phase term beyond the symplectic polygonal area, 
stationary phase evaluation of the multiple integral in \eref{Weylproduct} 
for all the variables $\{\x_{\nu}...\x_1\}$ collapses each side of the polygon in Fig.1, 
$\Vxi_{\nu}=0~...~\Vxi_1=0$, and hence the polygon itself, with $\x_{\nu}=...=\x_1=\x_0$.
The $\hbar$-dependent corrections can be calculated via a generalization \cite{Report} of the familiar
Groenewold product formula \cite{Groenewold}, but they will be only of second order in $\hbar$
for symmetrizations that guarantee a Hermitian product.

This simple classicality of the Weyl representation of a product of mechanical observables,
which is shared by the initial correlation,
\be
\C_0 \approx \int d\x_0~ A_{\nu}(\x_0)...A_1(\x_0) ~W(\x_0),
\label{C0}
\ee
may be destroyed as the observables evolve. Even so, the absence of high period oscillations 
in each of the Weyl symbols, $A_j(\x)$, still allows for stationary phase evaluation of \eref{evproduct}, analogous to that of \eref{Weylproduct}. 
Indeed, the expression \eref{compaction2} for the reduced action, $S'(\x_0)$, 
can now be reinterpreted as the composition of $2\nu+1$ transformations, but with zero action
for the unprimed variables. 
\footnote{Note the subtle difference with respect to the multiple integral \eref{compprop}:
There, one integrates over the primed variables, arguments for the Weyl symbols for $\opU_j$.
Here, we integrate first over the unprimed variables, which represent the mechanical observables, $\opA_j$.}
Then the stationary condition \eref{statcon} for each of these centres
collapses the corresponding side of the (reduced) dynamical polygon, 
that is $\Vxi_j(\x_j)=0$, as depicted in Fig 5. 
\begin{figure}[htb!]
\centering
\includegraphics[height=4cm]{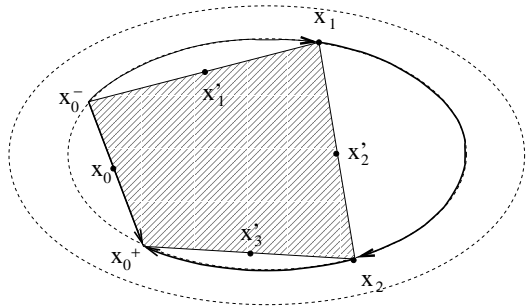}
\caption{Stationary phase integration over the unprimed reflection centres collapses the corresponding sides of the dynamical $(2\nu+1)$-sided polygon: These corners of the resulting $(\nu+1)$-sided polygon
now depend on the initial value, ${\x_0}^-$.}
\label{Figmechevol}
\end{figure}
Nonetheless, the {\it evolution sides}, ${\Vxi'}_j(\x'_j)$ are no longer zero, so there is generally a non-zero phase in the integrand that is constructed from the remaining $(\nu+1)$-sided polygon: 
Each unprimed variable, $\x_j$ is now a free corner, depending on the initial value ${\x_0}^-$,
instead of being the fixed centre of a side. 
In this way there results an enormous simplification of the correlation formula \eref{C4}:
\begin{eqnarray}
\C \approx \frac{1}{(\pi\hbar)^{\nu N}} \int d{\x_0}^{-}  &&               A_{\nu}(\x_{\nu}({\x_0}^{-}))...~A_1(\x_1({\x_0}^{-}))~W(\x_0({\x_0}^{-})) |\det(\Id + \M')|^{1/2} \nonumber \\
&&\exp\left[ \frac{i}{\hbar}({\mathbf S'}(\x_0({\x_0}^{-}))+ \hbar \pi \sigma)\right].
\label{C5}
\end{eqnarray}
Thus one no longer deals with a full family of compound canonical transformations. Instead of this, 
the polygonal trajectories are built for each initial value within a single canonical transformation without the intermediate reflections.  

In the limit of short times, $\nu$ sides of the remaining $(\nu+1)$-sided polygon shrink to zero.
In this limit the overall phase is zero, so that one retrieves \eref{C0} if $\nu$ is even.
This restriction on $\nu$ follows from the expression \eref{compmon} 
for the monodromy matrix in the amplitude of the compound propagator: As each
of the matrices $\M_j \rightarrow \mathbf I$, one obtains $\det(\Id + \M') \rightarrow 2^{2N}$
if $\nu$ is even, or zero if $\nu$ is odd (and hence a caustic). 
In the latter case, the Appendix obtains the correlation
as a single integral of the Weyl symbols for the observables weighed by the chord function
instead of the Wigner function.

The case of multiple Heisenberg evolution \eref{Heisenberg} also collapses
the SC phase, but for all time! The easiest way to see this is to propagate directly
the Weyl representation of each observable, $\opA(t)$:
\be
A(\x, t) = \int \frac{d\x_1d\x_2}{(\pi\hbar)^{2N}} [V(\x_1, t)]^* A(\x_2 +\x_1 - \x) V(\x_2, t)
\exp\left[\frac{i}{\hbar} \Delta _3(\x, \x_1, \x_2)\right],
\ee
so that the phase space point representing $\opA$ is placed at the corner of the triangle
oposite the point $\x$ where the evolved observable is evaluated. In \cite{Report} it is shown
that $A(\x,t)= A(\x'(t, \x))$, the classically evolved observable, for a metaplectic evolution,
i.e., for a quadratic driving Hamiltonian. This is also the correct semiclassical
approximation for a mechanical observable, resulting from stationary phase integration:
There is only a single trajectory traversed both forwards and backwards from the initial point, $\x$, 
and $\x_1=\x_2$ is its midpoint, so that there is complete phase cancellation.
In the case of the full correlation, the curved polygon in Fig 5 for the correlation collapses 
into a thin legged $\nu$-{\it spider} as shown in Fig 6.
Then one can merely obtain the evolved correlation from \eref{C0} with the 
classically evolved observables:
\be
\C \approx \int d\x_0~ A_{\nu}(\x'_{\nu}(\x_0))...A_1(\x'_1(\x_0)) ~W(\x_0).
\label{CHeis}
\ee
It is important to note that the collapse of the dynamical polygon into a spider concerns
exclusively the pairing of classical trajectories corresponding to the unitary operators within
the reduced compound operator. Thus there is no restriction on the Wigner function,
that is, the evolution is purely classical even for a highly oscillatory quantum Wigner function.
\begin{figure}[htb!]
\centering
\includegraphics[height=4cm]{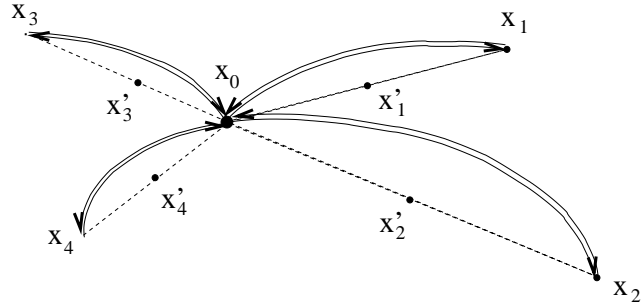}
\caption{ 
The dynamical polygon collapses for multiple Heisenberg evolution of mechanical observables, 
with each point $\x_j(\x_0, t)$ placed at the tip of a classical trajectory that is retraced with negative time. The {\it spider} shape arises when each observable is propagated by a different Hamiltonian. There is no phase factor in this essentially classical evolution 
for the correlation between mechanical observables. Any difference between forward 
and backward pair of Hamiltonians fattens the legs and these separate, 
even if the same pair propagates all the observables.}
\label{Figspider}
\end{figure}

\section{Discussion}

The elaborate theory underlying SC approximations for the multiple evolution of correlations 
between arbitrary numbers of quantum observables leads to deceptively simple results.
The approximate semiclassical scenario presented here becomes exact in the limit
where all the evolution operators are metaplectic, that is, generated by quadratic Hamiltonians.

Correlations are independent from the choice of representation that is employed in their calculation,
but the reliance on Weyl symbols and the Wigner function for the present theory 
leads to a rich phase space structure. The evolution kernel is identified
with a compound propagator, which is constructed by a dynamical polygon whose sides
are orbit segments that are combined into a compound classical trajectory 
and thence to a quantum phase.
Depending on the choice of symmetrization of the observables, an appropriate 
trigonometric function of the polygonal area will dephase the correlation integral. 

It is shown in the Appendix that the need to distinguish whether the number of observables 
is even or odd dissolves for short times and for the restricted class of {\it mechanical observables}, 
such that the Weyl representation of their product coincides with the corresponding smooth classical phase space function, except for corrections that are of first order in $\hbar$. The general rule is that independent Heisenberg evolution for each observable leads to \eref{CHeis} 
a classical expression of the evolving correlation, 
even if the Wigner function employed in their average has quantum oscillations 
that are separately though concurrently sampled by each observable. 

In contrast, if each observable does not follow its own Heisenberg evolution 
so that the intermediate evolution operators are not given by \eref{intermediate},
a phase factor will grow in time within the single phase space integral for the correlation.
This is generalizes the result for the fidelity (or the quantum Loschmidt echo) presented in {\bf I}, a special case within the present framework, that of a single {\it observable}, the identity operator, undergoing different forward and back evolutions.  In the language of section 5,
the dephasing grows with the symplectic area of a single slightly fattened spider leg, 
i.e. a curvilinear triangle. What about a repeated echo: 
On returning, one evolves again with the same pair of forward and back Hamiltonians? 
Then the initial value for the second traversal has changed, 
so that a new spider leg is drawn which is only initially close to the first leg, 
even though it is generated by the same pair of Hamiltonians 
(specially if they are chaotic). The correct symplectic area 
that determines the dephasing is then that of the full two legged spider, 
a curvilinear pentagon, instead of twice the area of the first triangular leg. 
For small times, the difference may be small, 
but the denominator for the phase is Planck's constant...

Whereas the simple {\it dephasing representation} of Van\'{i}\v{c}ek \cite{Vanicek} 
atributes the dephasing factor for the fidelity to a single classical trajectory, 
this was shown in \cite{ZamOA11} to result from a first order classical perturbation theory 
of an action for a pair of trajectories.
A further generalization in {\bf I} related the evolution of the expectation of a single observable 
to a pair of trajectories surrounding a translation or a reflection. Now we find that the only
price to pay for having more mechanical observables in a correlation is to add more segments 
to the corresponding compound classical trajectory. In all cases, the relevant trajectory is
completely specified by its initial value, i. e. the integration variable in the average.
Furthermore the general analysis of {\it Maslov phases} in {\bf II}, that is valid for all cases,
guarantees that initially, they are absent and it is only after a first caustic is crossed 
that extra phases need to be taken into account.

One should note that relaxing the restriction to mechanical observables does not necessarily
complicate matters. Observables may well be projectors, so that
the correlations become correlations between measurements. For instance,
one may measure the momentum (see \cite{Mil12} for examples), i.e. the projector, $\hat{B}_P$, 
rendered in the Weyl representation by the singular distribution, 
$B_P(\x)= \delta(p-P)$, which actually simplifies the multiple integral for a correlation.
More generally, positive operator valued measures (POVM, see \cite{Peres}) 
are also represented by phase space functions in the Weyl representation. 
A specially interesting case to study in this setting is that of a parity  
projection onto the subspace for either eigenvalue $\pm 1$
of the reflection operators. Reflections are observables as well as being unitary operators 
and their Weyl symbol is just a pointwise delta function. Their measurement was
proposed in \cite{Englert,LutterbachDav} and carried out experimentally in \cite{Bertet02}.
Thus, one may readily extend the present results 
beyond the restricted class of mechanical observables.

\appendix

\section{Mixed centre-chord propagation kernel}

Before any evolution takes place, that is, when all of the $t_j=0$ in \eref{C1},
the compound unitary evolution operator is just the product of reflections \eref{compu0}.
When $\nu$, the number of reflections is even, i.e. an odd number of observables,
the product is a translation, rather than a reflection and its trace is a delta function
in the Weyl representation \cite{Report}. Thus the zero time limit of  
${\rm tr}~ \widehat{\mathbf U}\{\x_0, \x_1, ..., \x_{\nu}\}= {\mathbf U'}(\x_0)$ is singular.

On the other hand, one may resort to the chord representation of the reduced compound unitary operator,
\begin{equation}
{\rm tr}~ \opU_{\nu+1}~\hat{R}_{\x_{\nu}}~\opU_{\nu}...~\hat{R}_{\x_1}\opU_1~\hat{T}_{-\Vxi_0}~
= ~ {\mathbf U'}\{\x_1, ..., \x_{\nu}\}(\Vxi_0),
\end{equation}
even though it is specified by its intermediate reflection centres,
so that the evolved product operator has its chord representation, 
\be
\fl \{A_{\nu}...A_1\}'(\Vxi_0) = \frac{1}{(\pi\hbar)^{\nu N}}\int d\x_{\nu}...d\x_1 ~ A_{\nu}(\x_{\nu})...A(\x_1) ~ 
{\mathbf U'}\{\x_1, ..., \x_{\nu}\}(\Vxi_0),
\ee
similarly to its Weyl representation \eref{Weylproduct}.
Thus, the general expression for the trace of a product in the chord representation \cite{Report}
supplies the correlation as 
\be
\C = \int d\Vxi_0~ \{A_{\nu}...A_1\}'(\Vxi_0)~ \chi(-\Vxi_0),
\label{CA}
\ee
whether or not the operators $\opA_j$ have evolved.  

In  effect this is a generalization
of the {\it correlation} for a single observable, i.e. its expectation in {\bf I}: 
The evolution was there atributed to the density operator, rather than to the single observable, but
here the opposite choice is preferable, because the observables may evolve independently.
One should note that, in the case of an even number of observables, it is the chord representation
that is singular initially. So there is no way around the need for different treatments 
depending on the parity.

To obtain the initial correlation for an odd number of observables, one needs the chord symbol
for an odd number of reflection operators. Following the relations in \cite{Report}, 
this is deduced to be
\be
\opR_{\x_{\nu}}...\opR_{\x_1} = \e^{\frac{i}{\hbar} \Delta _{\nu}(\x_1,..., \x_{\nu})} \opR_{\x_0},
\ee
where $\Delta_{\nu}$ is the area of the closed polygon with sides centred on $\{\x_1,..., \x_{\nu}\}$
as shown in Fig.1, even though the polygonal path generated by $\nu$ reflections 
from an arbitrary point is not closed. The segment that does close such a $(\nu+1)$-polygon
is necessarily centred on
\be
\x_0 = \x_{\nu} - \frac{1}{2}\Vxi_{\nu}(\x_1,..., \x_{\nu-1}),
\ee 
where, in its turn, $\Vxi_{\nu}$ is the open side of the $\nu$-polygon whose other sides
are centred on $\{\x_1,..., \x_{\nu-1}\}$. 
\footnote{If $\nu=3$, $\x_0$ is just the missing corner of the parallelogram with its other corners
at $\{\x_1, \x_2, \x_3\}$: The {\it inscribed polygon} defined in \cite{Report}. 
Inscribed polygons, for all odd $\nu$, satisfy special constraints.} 
It follows that
\be
\{R_{\x_{\nu}}...R_{\x_1}\}(\Vxi) 
=  \e^{\frac{i}{\hbar} \Delta _{\nu}} {\rm tr}~ \opR_{\x_0}~\opT_{-\Vxi}
= 2^{-N} \e^{\frac{i}{\hbar} \Delta _{\nu}(\x_1,..., \x_{\nu})} \e^{\frac{i}{\hbar} \x_0\wedge \Vxi} ,
\ee
so that the chord symbol for an arbitrary product of operators, 
each one specified by its Weyl symbol, is
\be
\{A_{\nu}...A_1\}(\Vxi_0) = \int \frac{d\x_{\nu}...d\x_1}{2^N (\pi\hbar)^{\nu N }} 
A_{\nu}(\x_{\nu})...A_1(\x_1)~ e^{\frac{i}{\hbar} \Delta _{\nu}(\x_1,..., \x_{\nu})}  \e^{\frac{i}{\hbar} \x_0\wedge \Vxi_0}.
\ee
Thus, according to \eref{CA}, the initial correlation in the case of an odd number of operators is
\begin{eqnarray}
\C_0 &&= \int \frac{d\x_{\nu}...d\x_1}{(\pi\hbar)^{\nu N }} 
A_{\nu}(\x_{\nu})...A_1(\x_1)~ e^{\frac{i}{\hbar} \Delta _{\nu}(\x_1,..., \x_{\nu})}  
\int \frac{d\Vxi_0}{2^N}~ \chi(-\Vxi_0)~ \e^{\frac{i}{\hbar} \x_0\wedge \Vxi_0}\nonumber\\
&&= \int \frac{d\x_{\nu}...d\x_1}{(\pi\hbar)^{(\nu-1) N }} 
A_{\nu}(\x_{\nu})...A_1(\x_1)~ e^{\frac{i}{\hbar} \Delta _{\nu}(\x_1,..., \x_{\nu})}  W(\x_0(\x_1,...\x_{\nu})). 
\end{eqnarray}

If all the operators, $\opA_j$, are now assumed to be mechanical observables with nearly
classical, non-oscillatory Weyl symbols, $A_j(\x_j)$, then the stationary phase evaluation
of the integral over $\x_j$ collapses the side of the polygon, 
which it centres, i.e. $\Vxi_j = 0$. Performing $(\nu-1)$ stationary phase integrations sequentially,
from $\x_{\nu}$ to $\x_2$, the polygon looses its sides, while the argument of the Wigner function 
in the remaining integrals becomes dependent on fewer variables. Finally, the last remaining integral
is just 
\be
\C_0 \approx \int d\x_1~ A_{\nu}(\x_1)...A_1(\x_1) ~W(\x_1),
\ee
which coincides with \eref{C0}, inspite of the number of observables being odd. This may now
be reinterpreted, such that the Weyl symbol for the product is approximately $A_{\nu}(\x)...A_1(\x)$,
so that the symbol for the evolved product is 
$\{\opA_{\nu}...\opA_1\}'(\x)\approx A'_{\nu}(\x)...A'_1(\x)$ and hence that \eref{CHeis}
holds approximately for the independent Heisenberg evolution of all operators, whether even or odd.

\section*{Acknowledgements}
We thank  Eduardo Zambrano and Raul Vallejos for stimulating discussions.
Partial financial support from the 
National Institute for Science and Technology--Quantum Information, 
FAPERJ and CNPq (Brazilian agencies) is gratefully acknowledged.

\end{document}